\documentclass[a4paper]{article}

\usepackage{mathrsfs}
\usepackage{amsmath, amsfonts}  
\usepackage[utf8]{inputenc} 
\usepackage[T1]{fontenc}    
\usepackage[hidelinks]{hyperref}       
\usepackage{url}            
\usepackage{booktabs}       
\usepackage{amsfonts}       
\usepackage{nicefrac}       
\usepackage{microtype}      
\usepackage{lipsum}
\usepackage{graphicx}
\usepackage[numbers, sort&compress]{natbib}
\usepackage[affil-it]{authblk}
\usepackage{fancyhdr}
\usepackage{color}

\fancypagestyle{firstpage}{\fancyhead[L]{\color{blue} 
}}

\author[1]{L. Antonucci}

\author[1]{A. Bonvalet}

\author[1]{X. Solinas}

\author[1]{M. Joffre}
\affil[1]{Laboratory for Optics and Biosciences, Ecole Polytechnique, 
CNRS, INSERM, Institut Polytechnique de Paris, 91120 Palaiseau, France}

\title{\bf Time delay distribution and laser stability in Arbitrary Detuning Asynchronous Optical Sampling.}

\date{}

\begin{document}
\maketitle


\begin{abstract}
  Arbitrary Detuning ASynchronous OPtical Sampling (ADASOPS) is an emerging technique for extending standard pump-probe experiments performed with two femtosecond lasers to multi-timescale experiments, which are of great interest for the study of complex systems. Although no specific requirements are needed for laser repetition rates, their ratio determines the achievable delay distribution and therefore is strongly related to the temporal resolution of the technique. We report a detailed theoretical analysis of measurement performances with respect to laser repetition rates and we validate our model with experimental data. In case of amplified laser systems, we demonstrate that achieved delays are inherently correlated to the time interval between amplified pulses, which affects the pulse energy and can generate artifacts. Nevertheless, a deep understanding of the origin of such artifacts allows to suggest several compensation strategies, either during data analysis, or at the conception of the experimental setup. Finally we present a new algorithm integrated into the ADASOPS device: by selecting pairs of probe pulses having the same elapsed time with respect to the previous pulse, it automatically compensates any effect of energy fluctuation. 
\end{abstract}

\thispagestyle{firstpage}

\section{Introduction}
Time resolved studies of complex systems can be extremely challenging due to their multifactorial response to stimulation, which results from a cascade of successive events.
A typical example of high interest is the internal dynamics of biological molecules. At short distances from the active site the process starts with elementary reactions, such as bond breaking, electron or proton transfer, etc., occurring in the femtosecond to picosecond timescale. In following stages, it evolves toward regained equilibrium through different, possibly more spatially extended, transition-state configurations that can take place over various timescales, going from nanoseconds to milliseconds and even seconds depending on the complexity of the molecule \cite{Ansari1985,Negrerie_12_acs,Nango2016,Sorigue2021, Aleksandrov2024}. Therefore, a comprehensive study of the dynamic response of a complex system requires the ability to perform an experiment over an extremely wide range of time domains.

Femtosecond pump-probe spectroscopy is a commonly used technique for time-resolved investigations \cite{Maiuri2020}. The ultrafast regime is especially accessible because the temporal resolution is only limited by the duration of the pump and probe pulses, which are typically of the order of tens of femtoseconds. The multi-timescale dimension can be achieved provided also the ability of scanning the two ultrashort pulses over a wide time range with high resolution, two characteristics that are difficult to attain with a single setup. 

At the state of the art, for most physical-chemistry laboratories, multi-timescale studies are mainly accessible using two different experiments: one for the investigation of the fast response - below a few nanoseconds - and a different one for the investigation of the longer lasting response of complex samples. Technologies allowing access to the whole dynamics with a single experiment, although much more attractive from the standpoint of repeatability and speed of data acquisition, are unaffordable for many experimentalists, because of the complexity and cost involved.
In this context, it would be extremely useful to have a technology for single-setup, multi-timescale experiments, requiring relatively affordable technical and financial resources. It would be even better if this technology could be put into operation by simply adding an electronic device to a standard pre-existent pump-probe experiment, which would potentially extend the possibility of multi-timescale approach to a large part of the ultrafast physical-chemistry community.

Typical femtosecond to picosecond resolution pump-probe setups are based on mechanical delay lines, for which maximum delays are of the order of a few nanoseconds and are obtained at the expense of pointing stability, focusing and scanning speed. An alternative technique, named ASynchronous OPtical Sampling (ASOPS), uses two femtosecond oscillators with slight repetition rate difference, so as to realize a stroboscopic sampling without any moving part and temporally limited only by the oscillator period \cite{Bartels_10_oe}. The drawback of ASOPS is that the resolution is related to the difference in repetition rates and its stability: for this reason, the method is most often applied to phase-locked oscillators and cannot be easily implemented on pre-existing laser systems. Also, ASOPS is applied to laser oscillators with typical periods of a few nanoseconds or more.
Multi-timescale experiments, aiming at reaching longer dynamics, generate pump and probe pulses with two separate femtosecond amplifiers. In this way the upper limit of the time delay is the amplifier period, which is typically in the millisecond region, and the scan step, realized by changing the oscillator pulse seeded into the amplifier, is limited to the oscillator period. However, by accepting increased experimental complexity, more precise adjustments can be achieved. On the one hand, the oscillators can be phase-locked so that fine tuning of the pump-probe delay is obtained by changing the relative phase between the oscillators \cite{Hamm_04_rsi, Ihalainen2007}. 
On the other hand it is also possible to seed the amplifiers with only one oscillator. In this case the unamplified beam is split in two and a conventional delay line is used to finely adjust the time delay between pump and probe pulses \cite{Mathes2015, Song2019}. 
In the last decade, more innovative strategies have been developed. Multiple probe approaches, based on new laser technologies with repetition rates greater than 100 kHz, demonstrated their multi-timescale capability  \cite{Greetham2012, Uriarte2022}. In the meantime, ADASOPS, a technique inspired by ASOPS but applied to lasers with Arbitrary frequency Detuning, proved capable of performing multi-timescale experiments with conventional laser systems \cite{Joffre_12_oe, Joffre_13_ol, Joffre_17_oe, Joffre_15_oe, Antonucci2020, Helbing2023}. 

\begin{figure}[!ht]
\centering
\includegraphics[scale=.42]{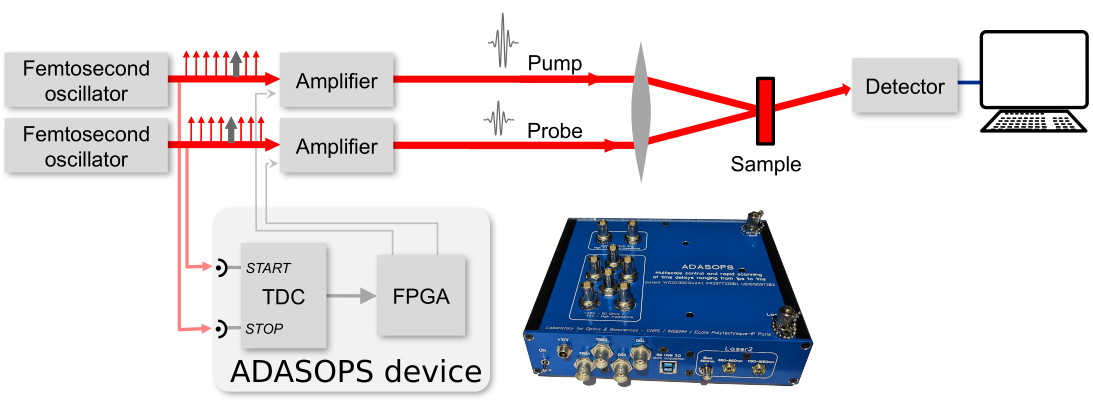}
\caption{Scheme of multi-timescale pump-probe experiment obtained by simply adding the ADASOPS device to a standard pump-probe setup based on amplified laser systems (kHz-ADASOPS). The device monitors the delay between oscillator pulses and triggers the amplifiers. In case of non-amplified systems (MHz-ADASOPS) the device only retrieves the delays.}
\label{setup_pump_probe}
\end{figure}

The ADASOPS implementation merely consists in adding an FPGA-driven electronic device to a standard pump-probe experiment, with very few optical additions to the setup (Figure \ref{setup_pump_probe}).
The fact that the ADASOPS method does not impose any condition on the laser repetition rates is a highly-valuable feature, which ensures that the method can be widely used, for example in laboratories where several lasers are already available, and enables pump and probe configurations to be varied with great flexibility. ADASOPS has already showed a variety of applications with standard systems \cite{Joffre_13_ol, Joffre_17_oe, Sorigue2021, Aleksandrov2024}.
In this context, it is of prime importance to understand whether different repetition frequencies have an effect on the measurement characteristics and, if so, to evaluate them in order to know in advance what performances can be expected from a specific setup.

In this article, we aim to show the effects of laser repetition frequencies on ADASOPS measurement performance.
In particular, we will focus on the distribution of delays and its effect on resolution and laser stability. First, we will study the delay distribution between oscillator pulses and between amplified pulses, find a mathematical description of how the resolution evolves, and discuss resulting energy fluctuations. We will validate our assertions by comparing simulations and experimental results. Eventually, we will discuss different methods for compensating for possible laser instabilities.

\section{The ADASOPS method}
\subsection{Asynchronous scanning}
\begin{figure}[!ht]
\centering
\includegraphics[scale=.6]{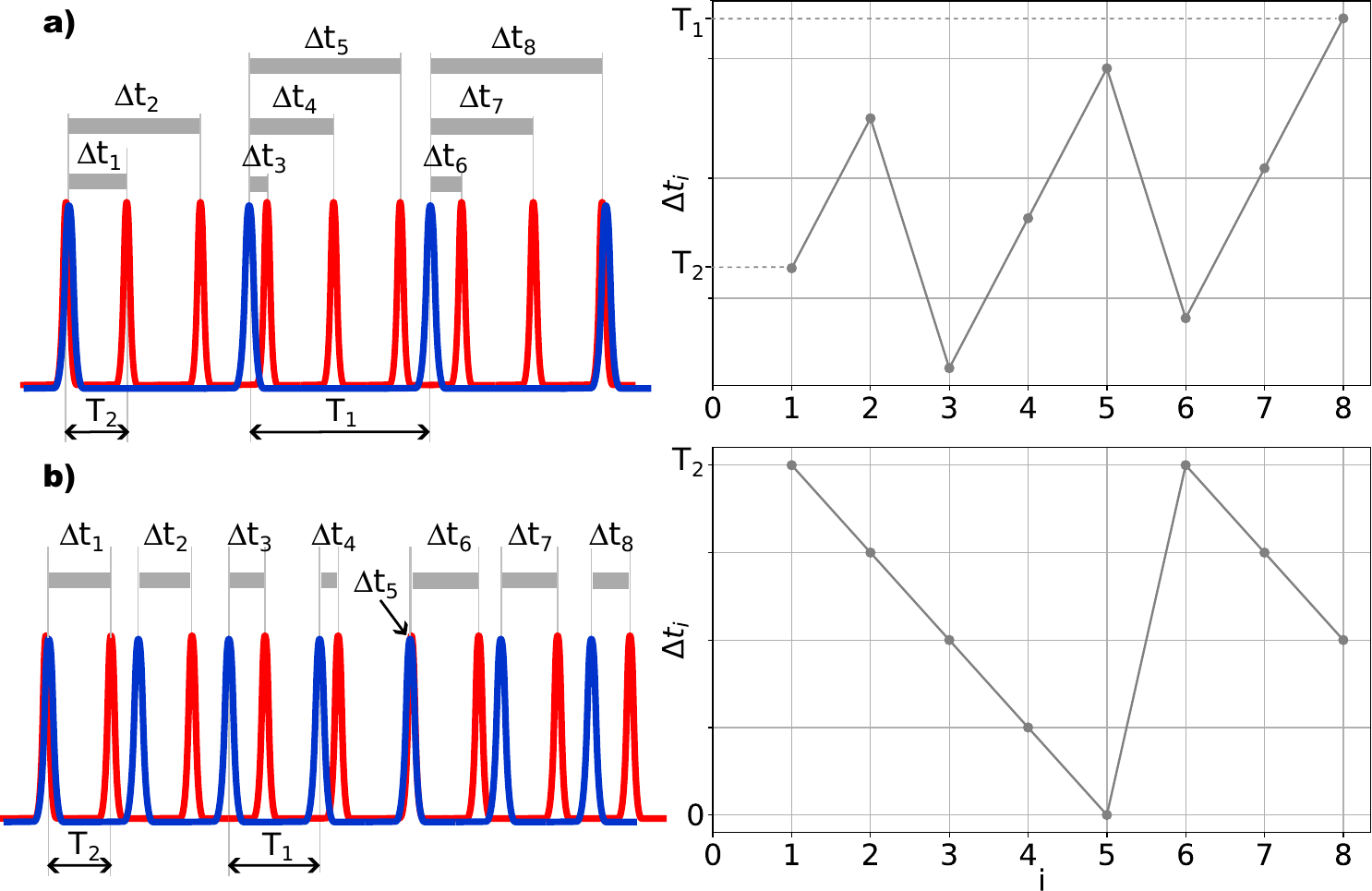}
\caption{Delay evolution due to asynchronous optical scanning in case of $f_1/f_2 = 3/8$ (a) and  $f_1/f_2 = 4/5$ (b). The left part of the figure shows the pulse-train diagram, with laser 1 in blue and laser 2 in red. The right part of the figure shows the time delay $\Delta t_i$ as a function of $i$, the pulse number of laser 2.} 
\label{principle}
\end{figure}

The ADASOPS method is a simple way of generating multi-timescale pump-probe delays. Its principle and the implementation details have been widely presented in previous publications \cite{Joffre_12_oe, Joffre_13_ol,Joffre_15_oe,Joffre_17_oe,Antonucci2020,Helbing2023}. But in order to be able to predict the delay evolution in all the different configurations that might arise by coupling two generic femtosecond laser systems, we focus first on the theoretical approach. As the acronym indicates, ADASOPS is based on ASOPS, except that, unlike the original version which generally applies to lasers of similar frequencies, the frequencies can now be very different, as shown in Fig.~\ref{principle}. 

Let us consider the delay evolution of two generic trains of femtosecond laser pulses with arbitrary repetition rates. We name by convention L$_1$, the laser with lower frequency $f_1$ and higher period $T_1=1/f_1$ (blue in the figure), and L$_2$ the one with higher frequency $f_2$ and lower period $T_2=1/f_2$ (red in the figure). We arbitrarily choose L$_1$ as reference laser and define for each pulse number $i$ from L$_2$ the time interval $\Delta t_i$ as its delay with respect to the previous L$_1$ pulse. Since the repetition frequencies are generally different, the delay will vary over time, increasing if $T_1>2 T_2$ (Fig.~\ref{principle}a), decreasing otherwise (Fig.~\ref{principle}b). In Fig.~\ref{principle}a,  $T_2 = 3/8\,\,T_1$: the delay increases linearly for the first two L$_2$ pulses until a new L$_1$ pulse arrives and becomes the new reference. The delay then drops close to zero and starts growing again until the next L$_1$ pulse and so on, evolving with a sawtooth behavior. In Fig.~\ref{principle}b, $T_2=4/5\,\,T_1$, so that every five L$_2$ pulses, three are univocally associated with a L$_1$ pulse and have linearly decreasing delay (i.e. $\Delta t_2$, $\Delta t_3$ and $\Delta t_4$) and two are associated with the same single L$_1$ pulse, one being in time coincidence and the other at $\Delta t_i=T_2$ (e.g. $\Delta t_5$ and $\Delta t_6$). Here again the delay has a sawtooth evolution, but with negative slope.

\subsection{Delay distribution between two oscillators}
In the context of pump-probe experiments, the general situations described above correspond to MHz-ADASOPS \cite{Joffre_12_oe, Joffre_13_ol}, i.e. the method where L$_1$ and L$_2$ are mode-locked oscillators delivering respectively pump and probe pulses. 
In this case, provided $f_1$ and $f_2$ are known and constant, it is possible to predict the delay evolution over time, more particularly what are all the achievable delays and how they are distributed over the whole time interval $[0, T_1[$, so as to estimate the expected pump-probe resolution. Incidentally, in case $L_2$ is used as pump and $L_1$ as probe, the time interval is reduced to $[0, T_2[$.

\begin{figure}[!ht]
\centering
\includegraphics[scale=.7]{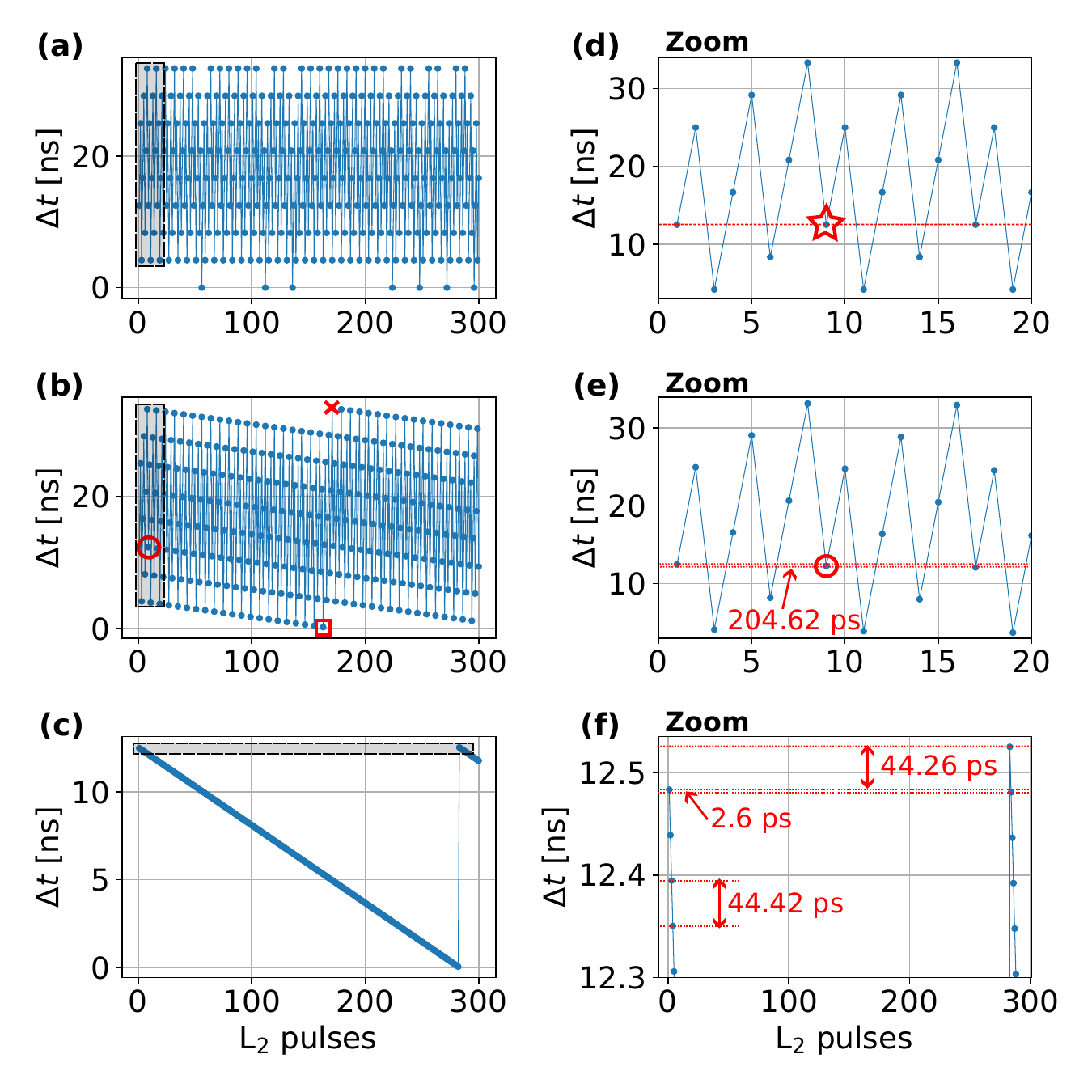}
\caption{Simulations of delay evolution between two oscillators with $f_1$=30,000,000~Hz and $f_2$=80,000,000~Hz (a and d), $f_1$ = 29,981,872~Hz and $f_2$ = 80,108,101~Hz (b and e), and $f_1$=79,825,000~Hz and $f_2$=80,108,124~Hz (c and f). Frequency values correspond respectively to Cases \#1, \#2 and \#3 as defined in Table \ref{TableOsc}. Graphs on the right provide a close-up of the shaded gray areas outlined on the left graphs.}
\label{oscillators}
\end{figure}

The diagram of Fig.~\ref{oscillators} shows the delay evolution in three different situations summarized in Table~\ref{TableOsc}. Fig.~\ref{oscillators}a represents a very simplified case where $f_1$ et $f_2$ are strictly equal to 30~MHz and 80~MHz (Case \#1 in Table~\ref{TableOsc}). As in Fig.~\ref{principle}a, their ratio is equal to the fraction $3/8$, and therefore the delay has a sawtooth evolution with positive slope. Every 2 to 3 pulses there is a new folding of the delay into the range $[0,T_1[$ and the maximum number of divisions of the total time interval $T_1$ is 8. The best resolution is $T_1/8 \approx 4. 17$~ns and it is reached after 8 pulses. From the $9^{th}$ pulse (star in the figure) the delay always returns to a value obtained previously.

Fig.~\ref{oscillators}b is an experimental situation close to (a), where $f_1=29,981,872$ Hz and $f_2=80,108,101$ Hz (Case \#2 in Table~\ref{TableOsc}). These are the frequencies actually measured for respectively the integrated fiber oscillator of a Quantronix Integra C system and a Coherent Vitara-S Titanium:Sapphire oscillator. Note that we round the frequencies to the nearest integer, because the Allan deviation of the two oscillators is measured to be of the order of 2~Hz for an observation time of 1~sec and decreases for longer times. As 80,108,101 happens to be a prime number, the frequency ratio $f_1/f_2$ is already a reduced fraction. This yields, analogously to the previous reasoning, a final resolution of 0.42~fs after 80,108,101 laser pulses. 
In short, the denominator of the reduced fraction associated to the ratio between oscillator frequencies determines the minimum achievable delay resolution.

The mathematical concept of continued fraction provides more insight into the behavior of the delay evolution. As shown in Eq.~\ref{eqn:continuedFraction}, every rational number $N/D$ can be expressed by a continued fraction of $n-1$ integer divisions, where $a_0, \, a_1, \,...,\, a_n$ are integer numbers:

\begin{equation}
  \frac{N}{D} = a_0 + \frac{1}{a_1+ \frac{1}{...+\frac{1}{a_i+\frac{1}{...+\frac{1}{a_n}}}}} = [a_0; a_1, ..., a_n]. \label{eqn:continuedFraction}
\end{equation}

The series of continued fractions, calculated from $a_0$ and considering each time an additional integer, is a series of rational numbers approximating $N/D$ more and more precisely. We call [$a_0;a_1$, ..., $a_m$], where $0 \leq m \leq n$, the $m^{th}$ convergent to [$a_0$; ..., $a_n$]. The convergents with $m=2j$, with $j \in \mathbb{N}$ and $j<n/2$, are called even convergents, while those with $m=2j+1$ are called odd convergents. It can be shown that odd convergents are greater than $N/D$ while even convergents are smaller until convergence \cite{Krishnan16}.

Applied to $f_1$ and $f_2$, Euler's algorithm allows the calculation of the integers $a_i$ associated with the continued fraction corresponding to $f_1/f_2$. In the previous case for example, the continued fraction of 3/8 is:

\[\frac{3}{8} =\frac{1}{2+\frac{1}{1+\frac{1}{2}}} = [0; 2, 1,2].\]

\begin{table}
  \caption{Summary for a two-oscillator system}
  \label{TableOsc}
  \includegraphics[width=\linewidth]{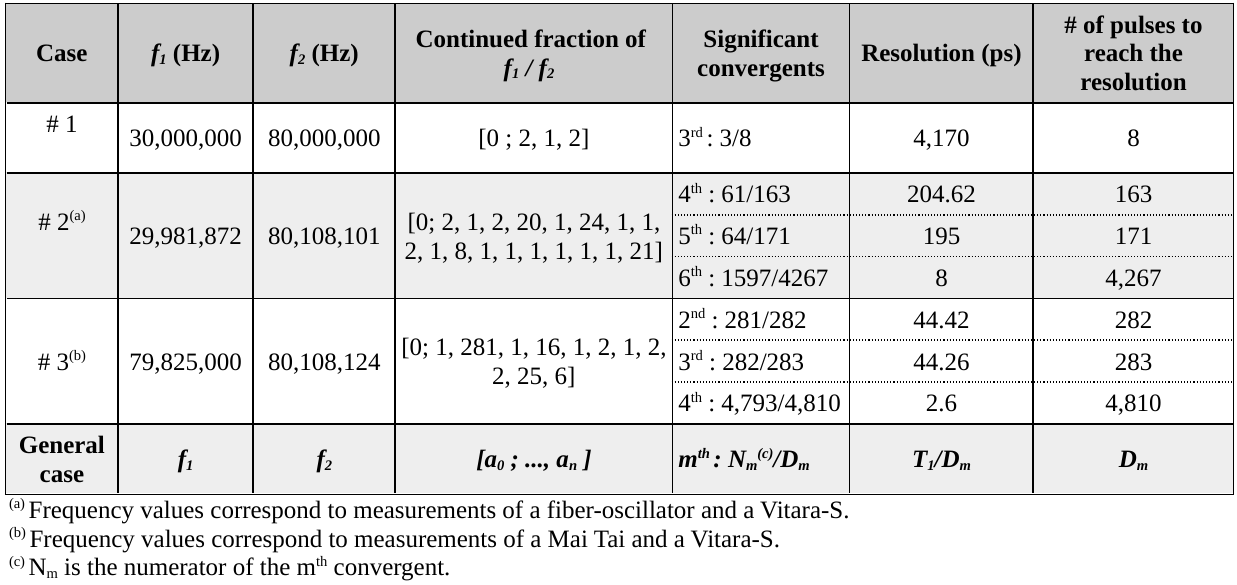}
\end{table}

In Case \#2 we obtain instead: 29,981,872/80,108,101 = [0; 2, 1, 2, 20, 1, 24, 1, 1, 2, 1, 8, 1, 1, 1, 1, 1, 1, 21]. The latter continued fraction has many more integer divisions than the one representing 3/8. Nevertheless, the first four integers are the same, therefore 3/8 is the third convergent of 29,981,872/80,108,101 and overestimates it. This explains the similarity of the delay evolution on a short time scale, as shown in Fig.\ref{oscillators} (d) and (e). On a longer time scale, shown in (a) and (b), the delays are aligned in both cases along 8 parallel lines. However, unlike (a) which exhibits horizontal lines, the lines shown in (b) exhibit a negative slope because each delay of (b) is less than each corresponding delay of (a). In order to quantitatively describe this pattern we need to calculate further convergents:

\begin{align}
  \frac{29981872}{80108101} &\approx [0; 2, 1, 2, 20]=\frac{61}{163} 
  \label{conv4}
  \\
&\approx [0; 2, 1, 2, 20,1]= \frac{64}{171} 
  \label{conv5}
\\
&\approx [0; 2, 1, 2, 20,1,24]= \frac{1597}{4267} 
  \label{conv6}
\end{align}

The fourth convergent 61/163 (Eq.~\ref{conv4}) indicates that from the $9^{th}$ pulse (red circle in Fig.~\ref{oscillators}b and e) the delay will not fall over a previously achieved value, but it will be slightly smaller, as compared with the closest delay among the previous ones, at every new pulse, until the $163^{rd}$ pulse (square in Fig.~\ref{oscillators}b) when the whole interval will be divided in 163 intervals. The shift can be approximated as $T_1/163 = 204.62$~ps. At pulse 164 (cross in the drawing), again the delay will shift, but now upwards, by an amount of approximately $T_1/171=195.04$~ps - where 171 is the denominator of the fifth convergent (Eq.~\ref{conv5}) - until pulse number 171 (red cross in  the  drawing) and the whole interval will be divided in 171 parts. From pulse  172,  the  delay  will continue shifting,  but now downwards and by an amount of approximately $T1/4,267 = 7.82$~ps, according to Eq.~\ref{conv6}, and so until the $4,267^{th}$ pulse. Using the following convergents we can carry on analogous observations about how the delay pattern evolves as pulses accumulate, and therefore how the resolution improves.

In summary, if $i$ is the number of L$_2$ pulses, for each $i=D_m$, where $D_m$ is the denominator of the $m^{th}$ convergent, the interval between delays is approx. $T_1/D_m$. For $D_{m-1}\leq i\leq D_m$, the $m-1$ intervals into which $T_1$ is divided for $i=D_{m-1}$ will be progressively reduced and the delays already achieved will be flanked by new delays at a distance ca. $T_1/D_m$ on the side of the longest delays if $m$ is odd or on the side of the shortest delays if $m$ is even. All approximations improve with increasing $m$ until they become exact for $m=n$.

Fig.~\ref{oscillators}c) illustrates a different situation, identified as Case \#3 in Table~\ref{TableOsc}, where $f_1=79,825,000$ and  $f_2=80,108,124$ are the frequencies measured for two Titanium:Sapphire oscillators: Mai Tai (Spectra-Physics) and Vitara-S (Coherent). Here the reduced fraction is 19,956,250/20,027,031, associated with the continued fraction [0; 1, 281, 1, 16, 1, 2, 1, 2, 2, 25, 6]. The first meaningful convergent is [0; 1, 281]=281/282, the second is [0; 1, 281, 1]=282/283 and the third [0; 1, 281, 1, 16]=4,793/4,810. We deduce that the delay decreases continuously for the first 282 pulses, analogously to Fig.~\ref{principle}b, slightly shifting with respect to the previous one by $T_1/282 = 44.42$~ps, as a first approximation. At the 283$^{rd}$ pulse the L$_2$ pulse train overtakes L$_1$ and the delay folds up at roughly $T_1$, shifted up compared to the $1^{st}$ delay by $T_1/283 = 44.26$~ps. Then it continues decreasing, generating new delays that are shifted down by $T_1/4810 = 2.60$~ps with respect to the ones in the first sawtooth, and so on.

To generalize, given two laser oscillators, the delay evolution between their pulses can be known by calculating the ratio between their frequencies: while the reduced fraction gives the maximum achievable resolution and the number of pulses required to achieve it, each integer of the continued fraction represents a step forward in improving the resolution. Let us now compare these findings with actual experimental data.

\begin{figure}[!ht]
\centering
\includegraphics[scale=.9]{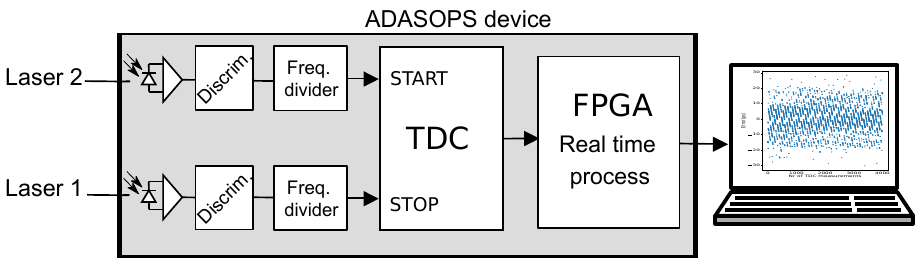}
\caption{Setup for TDC measurement of time delay between pulses from fiber-based oscillator (Laser 1) and Vitara-S (Laser 2).}
\label{setup}
\end{figure}

In order to validate our simulations with experimental data, we used the oscillators corresponding to Case \#2 in Table~\ref{TableOsc} - the Integra C fiber-based oscillator (L$_1$) and the Vitara-S (L$_2$) - with the electronic setup for delay measurement shown in Fig.~\ref{setup} \cite{Antonucci2020}. The experimental measurements were made using two amplified photodetectors (HFD3180-203, Finisar), two leading edge discriminators, two frequency dividers, and a TDC (Time to Digital Converter, THS788 Texas Instruments) for measuring the interval between pairs of L$_1$ and L$_2$ pulses at a maximum frequency of 10 MHz with ca. 10~ps accuracy and 13~ps precision.

\begin{figure}[!ht]
\centering
\includegraphics[scale=.42]{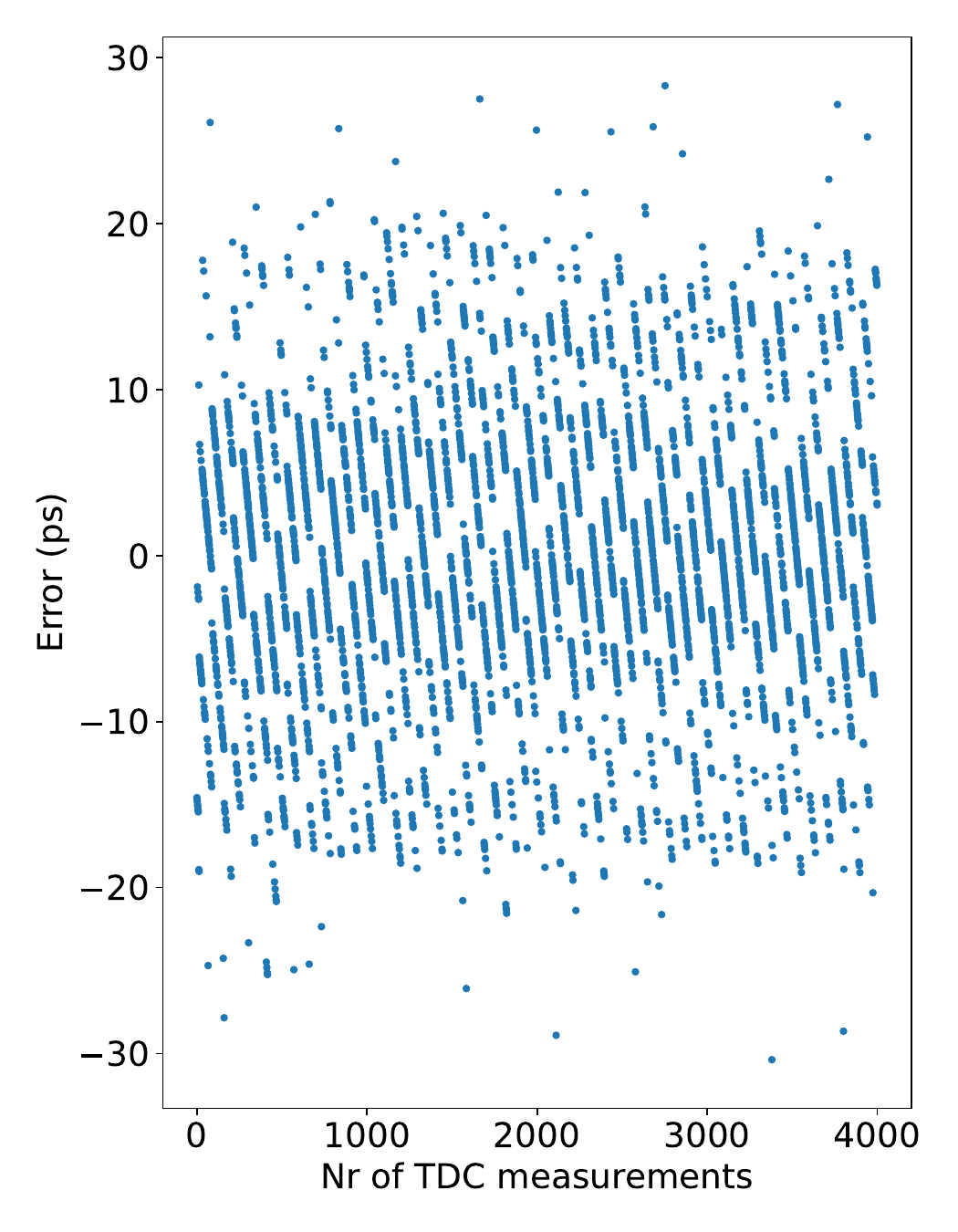}
\caption{Error between TDC measurements and theoretically calculated delays for fiber-based oscillator ($f_1$) and Vitara-S ($f_2$).}
\label{error_data_simu}
\end{figure}

Figure~\ref{error_data_simu} shows the error between 4000 delays measured by the TDC and the respective values expected according to a simulation assuming a perfectly stable repetition frequency.  

The pattern in the error plot is due to error steps of 13~ps or multiples of this number, and is due to the internal workings of the TDC. The error dispersion is  9~ps RMS (Root Mean Square) and can be entirely attributed to the electronic delay measurement chain. We can conclude that our model is consistent with the experiment, provided that the frequencies are stable. In fact the absence of significant deviation of the average error with respect to zero implies that the laser frequencies are stable within the time interval of the plot. But if we extended the plot to larger time scale a deviation would appear.

The ADASOPS method makes a more accurate estimation of the delay evolution by regularly fitting packets of TDC measurements using linear regression \cite{Antonucci2020}, i.e. assuming that the laser frequencies stay constant. The choice of packet size then must be properly chosen according to the lasers stability in term of repetition rate. For instance in the case of Vitara-S and the fiber-based oscillator we average 4000 TDC measurements so that we can expect to estimate the delays with a resolution of $9/\sqrt{4000}\approx0.14$~ps RMS at best.

Summarizing, the delay evolution between two femtosecond oscillators is fully determined by the ratio between their frequencies: the reduced fraction makes possible to calculate the best achievable delay resolution, while the expression as continued fraction gives the evolution of the resolution as the number of pulses increases. For a number of pulses increasing between two denominators of consecutive convergents the resolution will regularly improve and will have a faster improvement at each new convergent. A quantitative description will be presented in the context of amplified pulse selection in section \ref{delay_in_shoot_win}.

\subsection{Delay distribution between two amplified systems}
\label{ampli_delays_distribution}

\subsubsection{Pulse selection in kHz-ADASOPS}

\begin{figure}[!ht]
\centering
\includegraphics[scale=.6]{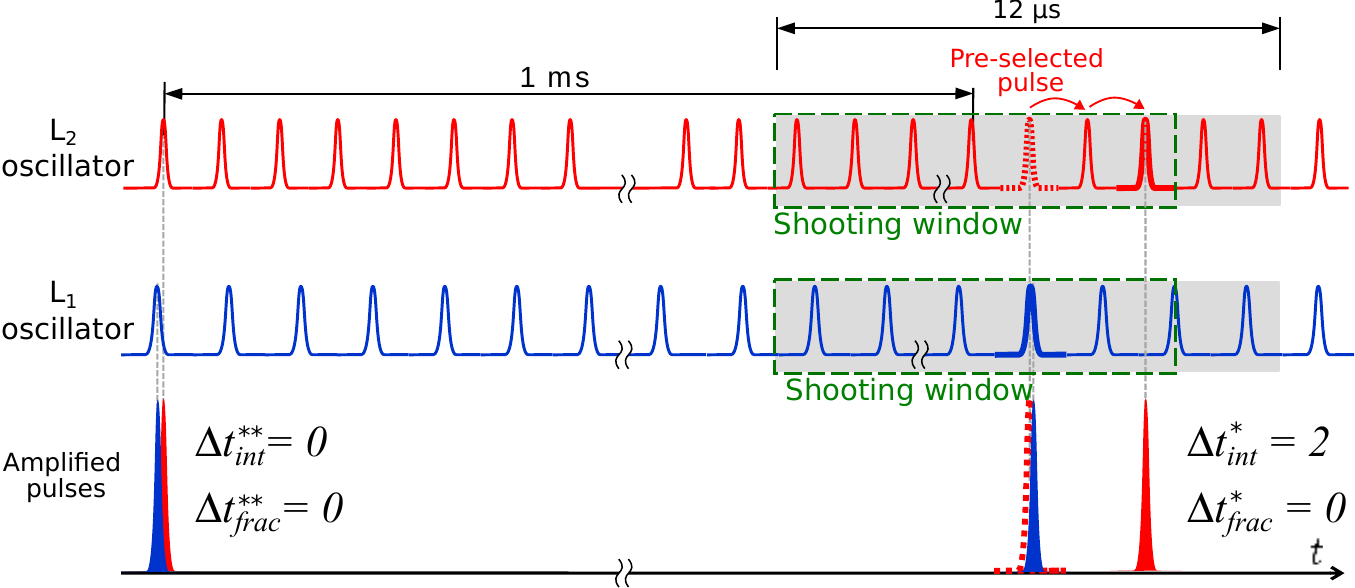}
\caption{Scheme of selection process in the case of $f_{trig}$ = 1 kHz. The target delay is $\Delta t^* = 2\cdot T_2$, while the previous delay was $\Delta t^{**} = 0$. The L$_2$ shooting window has been reduced of $\Delta N = \Delta t_{int}^*-\Delta t_{int}^{**}=2$ pulses. The L$_1$ window is also reduced as a consequence. }
\label{trigger_scheme}
\end{figure}

The knowledge of delay distribution between two oscillators is very useful also in kHz-ADASOPS, i.e. with amplified laser systems. Unlike MHz-ADASOPS, here there is an active control of pulse delay by the ADASOPS device as illustrated in Fig.~\ref{setup_pump_probe}. The device suitably triggers the amplifiers in order to amplify among the oscillator pulses a pair approximating at best the user-given target  \cite{Joffre_17_oe}. In Fig.~\ref{trigger_scheme} is depicted a scheme illustrating the selection process.

Let us refer to the amplifier frequency as $f_{trig}$. We conveniently express a delay $\Delta t\in[0, 1/f_{trig}[$ as the sum of two components, the integer part $\Delta t_{int}$ and the fractional part $\Delta t_{frac}$ defined as follows:

\begin{align}
\Delta t_{int} &= \biggl\lfloor\frac{\Delta t}{T_2}\biggl\rfloor \\
\Delta t_{frac} &= \Delta t - \Delta t_{int} ,
\end{align}
where the $\lfloor...\rfloor$ symbol stands for the floor operator.

When aiming a target delay $\Delta t^* = \Delta t_{int}^* + \Delta t_{frac}^*$, for each oscillator we call shooting window a set of $N$ candidate pulses centered $\lfloor1/f_{trig}\rfloor$ pulses further away from the last amplified pulse. $N$ is chosen so as not to vary the laser frequency by more than $\pm$0.6\% and thus  ensure the proper functioning of the amplifier. A typical value of $N$ for 100-MHz oscillators seeding 1-kHz amplifiers is thus 1200.

The choice of the amplified pulse pair is made by neglecting at first the integer part and identifying a pulse pair within the shooting window with delay having the fractional part closest to $\Delta t_{frac}^*$. Such a pair has the same integer part as the previous amplified pair (addressed as $\Delta t_{int}^{**}$ ). This selects the L$_1$ pulse. Taking now account of the integer part, the L$_2$ pulse will be the one shifted by a number of pulses $\Delta N = \Delta t_{int}^*-\Delta t_{int}^{**}$.  In order to keep the frequency change smaller than 0.6\%, $N$ must be decreased by $|\Delta N|$ at the beginning or at the end of the shooting window according to the sign of $\Delta N$. 
As a consequence, when a large delay step is requested, the aiming precision is reduced according to the width of the step. In particular, for a 12~µs step the shooting window is reduced to a single pulse, that is the delay of the amplified pair of pulses will have an uncertainty of $T_2$. This is not a major problem though because for high delays the aiming precision is less crucial and it is always possible to repeat the delay and get right back to the best precision. However, for delay steps higher than 12~µs, which would demand a frequency variation greater than 0.6\%, it is necessary to approach the target delay by several pairs of pulses with intermediate delays in order to distribute the frequency variation over them. For example, if we want to change the delay by 120~µs, the ADASOPS device will get there in 10 steps of 12~µs. This is problematic only when one wants to perform rapid scanning \cite{Joffre_17_oe}, in which case there is a compromise to make between scanning frequency and delay span.

In conclusion, in kHz-ADASOPS the electronic device triggers the amplifiers to select a pulse pair having a delay as close as possible to the target. The choice is made among a set of oscillator pulses centered one amplifier-period away from the previous amplified pulse (shooting window). The error between the target and the achieved delay depends on the available delays, thus is related to the oscillator frequencies, the amplifier frequency and the amplifier tolerance to frequency variation, which defines the window width.

\subsubsection{Delay distribution in the shooting window}
\label{delay_in_shoot_win}

\begin{figure}[!ht]
\centering
\includegraphics[scale=.5]{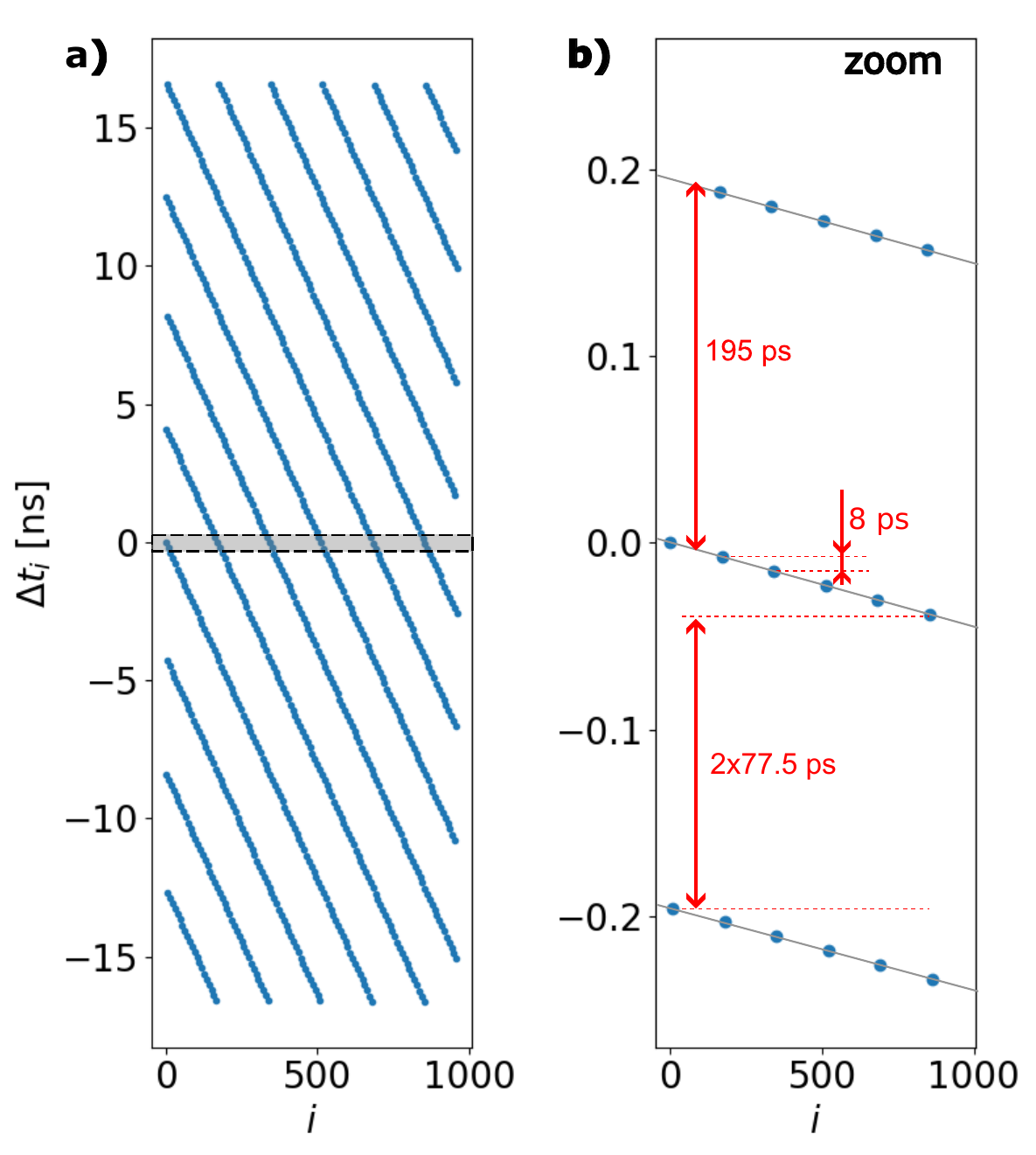}
\caption{(a) Delay distribution in the shooting window. (b) Zoom of the delay distribution plotted in (a) with respect to the gray box. The time intervals between delays depend on the ratio between oscillator frequencies and can be calculated using the series of convergents that approximate $f_1/f_2$.}
\label{shooting_window}
\end{figure}

Based on the previous discussion about the distribution of delays between oscillators, we have a very precise knowledge of the available delays in the shooting window. Fig.~\ref{shooting_window} shows the example of amplified laser systems with oscillator frequencies $f_1$=29,981,872~Hz and $f_2$=80,108,101~Hz (Case \#2) and amplifier frequency $f_{trig}=1$~kHz (amplifier period $T_{trig} = 1/f_{trig}=1$~ms). The number of pulses into the L$_2$ shooting window is $N=961$.

According to the previous calculations of continued fraction, in particular the fifth convergent (Eq.~\ref{conv5}), we know that the first 171 pulses of the shooting window divide $T_1$ into 171 different delays separated by 195~ps. Also, since the even convergent overestimates the real frequency ratio, from pulse number 172 to 801, each pulse will generate a new delay smaller than the closest previous delay by, according to Eq.~\ref{conv6}, $T_1/4267\approx 8$~ps, i.e. drawing a delay pattern with negative slope. This is illustrated in Fig.~\ref{shooting_window}b, which is an enlargement of the gray area of Fig.~\ref{shooting_window}a, and where the grey lines are plotted over delays separated by 171 pulses. Hereafter we will continue to refer to these delay alignments as “the gray lines”. Eventually, at the end of the window, for each of the 171 delays separated by 195~ps there will be at least $\lfloor (961-171)/171\rfloor=4$ other neighbouring delays separated by 8 ps (e.g. the upper gray line in Fig.~\ref{shooting_window}b), and for $961 \bmod 171 = 106$ delays there will be 5 neighbouring delays (e.g. lower and middle gray lines in Fig.~\ref{shooting_window}b).

This leads to two important consequences. First, for each shooting window the delay of the amplified pair can have a maximum error of $\pm(195-5\times8)/2=\pm77.5$~ps with respect to the target, or $\pm8/2=\pm4$~ps in the lucky but rare cases where the target delay falls onto one of the gray lines. Second, because of the slope of the gray lines, for each amplified pulse there is a correlation between the realized delay and the elapsed time $\mathscr{T}$ from the previous amplified pulse. In particular, the correlation is inverted compared to the slope, e.g. in the case of the figure a delay smaller than the target means $\mathscr{T}<1$~ms and vice-versa.

Generalizing: given a shooting window composed of $N$ pulses, with $D_{m}\leq N \leq D_{m+1}$, where $D_{m}$ and $D_{m+1}$ are the denominators of the $m^{th}$ and $(m+1)^{th}$ convergents of the continued fraction associated to $f_1/f_2$, we can assert:

\begin{itemize}
    \item the minimum and maximum number of intervals on the gray lines are respectively
        \begin{equation}
            k_{min}=\biggl\lfloor \frac{N-D_{m}}{D_{m}} \biggl\rfloor
          \label{k_min}
        \end{equation}
        and
        \begin{equation}
            k_{max}=\biggl\lceil \frac{N-D_{{m}}}{D_{m}} \biggl\rceil
          \label{k_max}
        \end{equation}
    where the $\lceil...\rceil$ symbol stands for the ceiling operator.
    \item The minimum and maximum number of delays on the gray lines are respectively
        \begin{equation}
            p_{min} = k_{min}+1
          \label{n_min}
        \end{equation}
        and
        \begin{equation}
            p_{max} = k_{max}+1.
          \label{n_max}
        \end{equation}
    \item the general peak-to-peak aiming resolution is:
        \begin{equation}
            T_1 \left(  \frac{1}{D_{m}} -  \frac{k_{min}}{D_{m+1}} \right)
            \le \frac{T_1}{D_m}.
            \label{aiming_res}
        \end{equation}
    \item the peak-to-peak aiming resolution for the particular cases where the delay target is on a gray line is:
        \begin{equation}
            \frac{T_1}{D_{m+1}}
            \label{pick_res}
        \end{equation}
    \item If $m$ is odd, the gray lines will have a negative slope and the correlation delay vs. $\mathscr{T}$ is direct. If $m$ is even, the gray lines will have a positive slope and the correlation delay vs. $\mathscr{T}$ is inverse.
\end{itemize}

Lastly, at any new shooting window the delay distribution will shift by:
\begin{equation}
            \biggl(  \biggl\lfloor\frac{T_{trig}}{T}\biggl\rfloor -  \frac{N}{2} \biggl)T_2\bmod T_1\bmod \frac{T_1}{D_m} \pm p\frac{T_1}{D_{m+1}},
            \label{histo_shift}
        \end{equation}
where $\lfloor T_{trig}/T\rfloor -  N/2$ is the number of oscillator pulses from the amplified pulse to the first pulse of the following shooting window ($T$ is $T_1$ or $T_2$ depending on the laser at issue). Mod~$T_1$ takes into account the fact that we are discussing the fractional part of the delay, $\bmod\frac{T_1}{D_m}$ considers the periodical pattern of the delay distribution and $\pm p\frac{T_1}{D_{m+1}}$, with $p\in\mathbb{N}$ and $0\leq p \leq n_{max}$, is the uncertainty depending on which one of the N pulses was selected for amplification in the previous shooting window. For example in the case of Fig.~\ref{shooting_window}, considering zero target delay, the second $L_2$ shooting window shifts by 145~ps compared to the first one.
In Table \ref{TableAmpli} are reported the main results and the general case discussed in the present section.

\begin{table}
  \caption{Summary for a two-amplifier system (kHz ADASOPS) based on a fiber oscillator and a Vitara-S oscillator}
  \label{TableAmpli}
  \includegraphics[width=\linewidth]{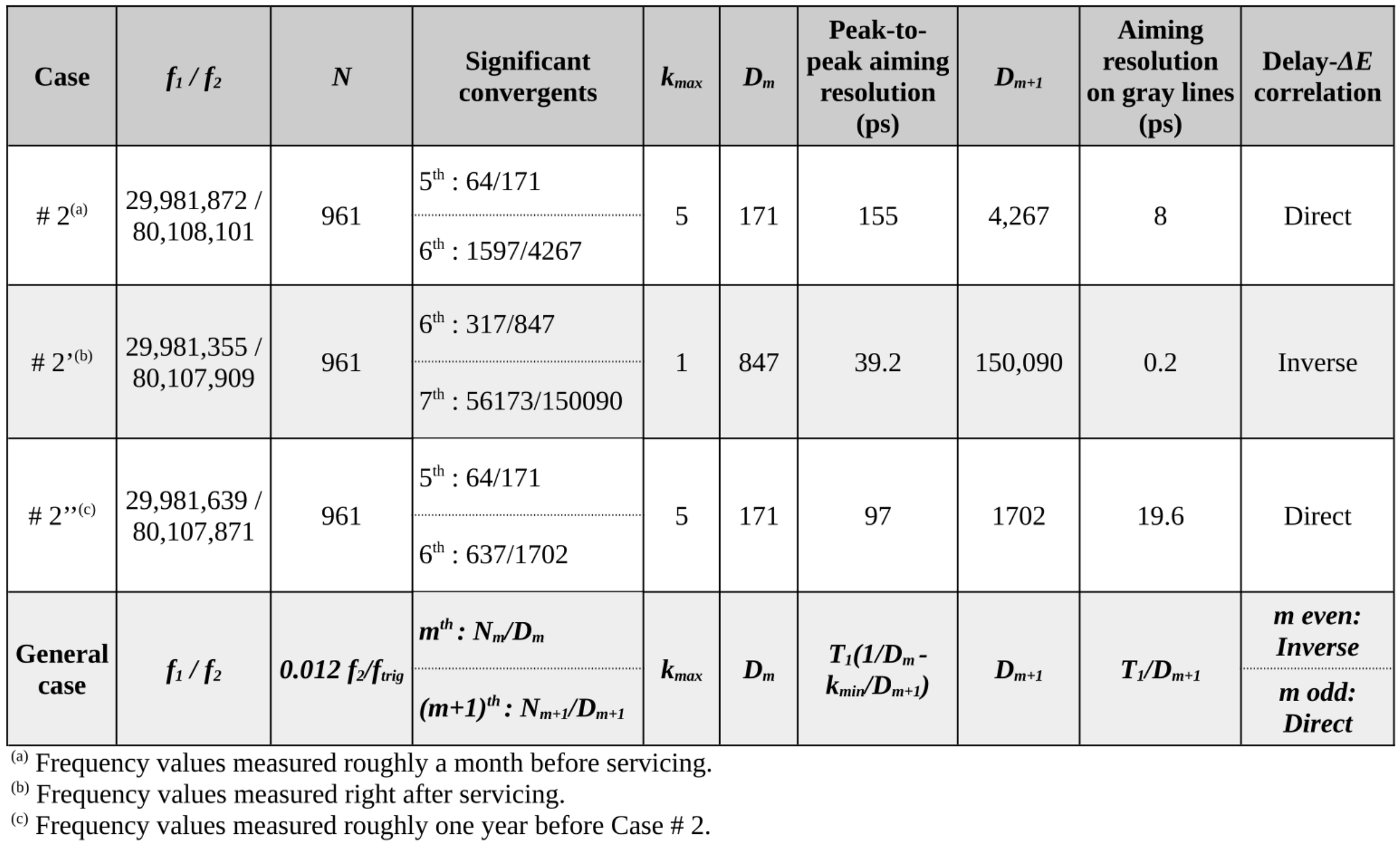}
\end{table}

\subsubsection{Delay distribution of amplified pulses}
\label{simu:data}

\begin{figure}[!ht]
\centering
\includegraphics[scale=.5]{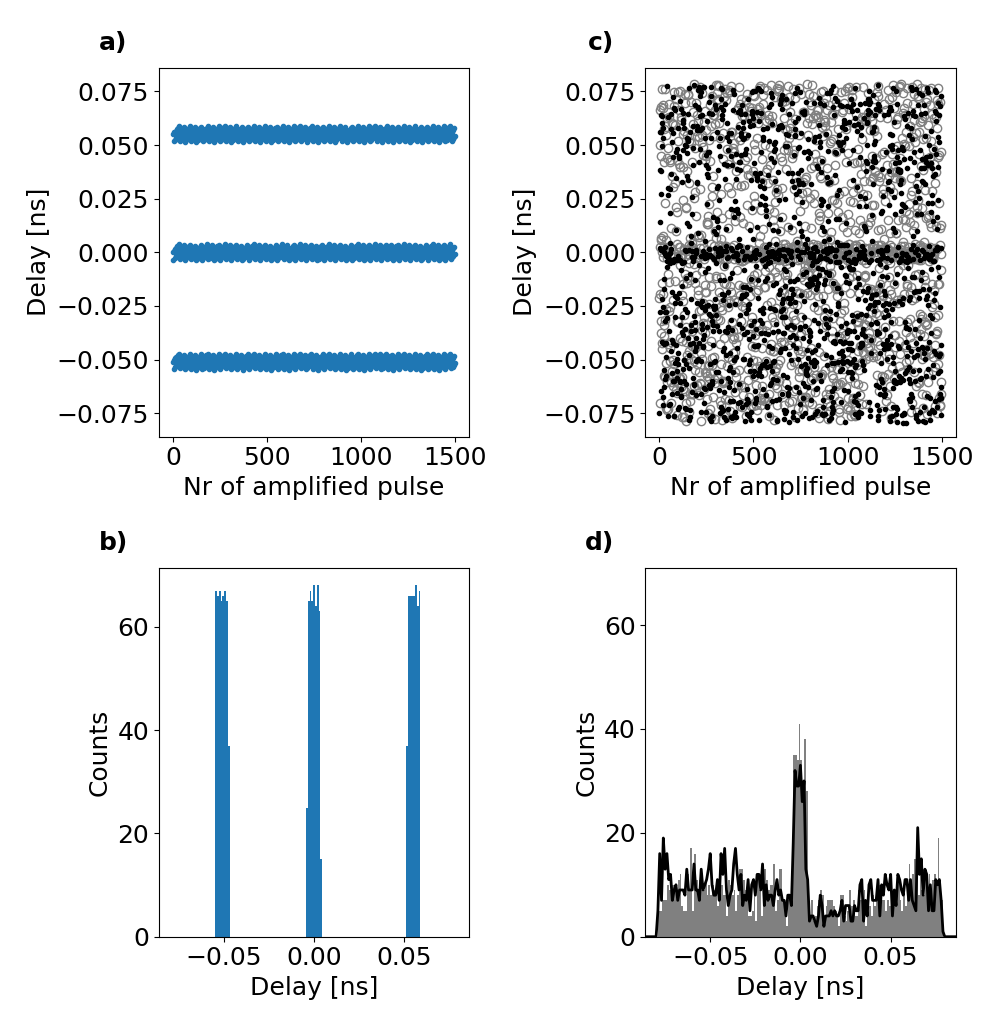}
\caption{Delay distribution for 1500 amplified pulse pairs for two laser systems having $f_1$=29,981,872~Hz, $f_2$=80,108,101~Hz and amplifier frequency $f_{trig}=1$~kHz: (a) Simulation of delay vs. pulse number without noise, (b) histogram of delay distribution without noise. In (c) and (d) the gray plots show the same quantities as in (a) and (b) but taking noise into account, while the black plots are experimental data.}
\label{fig_ampli}
\end{figure}

Taking into account all the considerations discussed so far, we carried out some simulations to predict the delays of amplified pulses selected according to the method described above, while targeting zero delay. In Fig.~\ref{fig_ampli}a and \ref{fig_ampli}b are reported the results calculated for two laser systems having $f_1$=29,981,872~Hz, $f_2$=80,108,101~Hz (Case \#2) and amplifier frequency $f_{trig}=1$~kHz, considering a total time of 1.5 seconds (1500 pulses).

\begin{figure}[!ht]
\centering
\includegraphics[scale=.4]{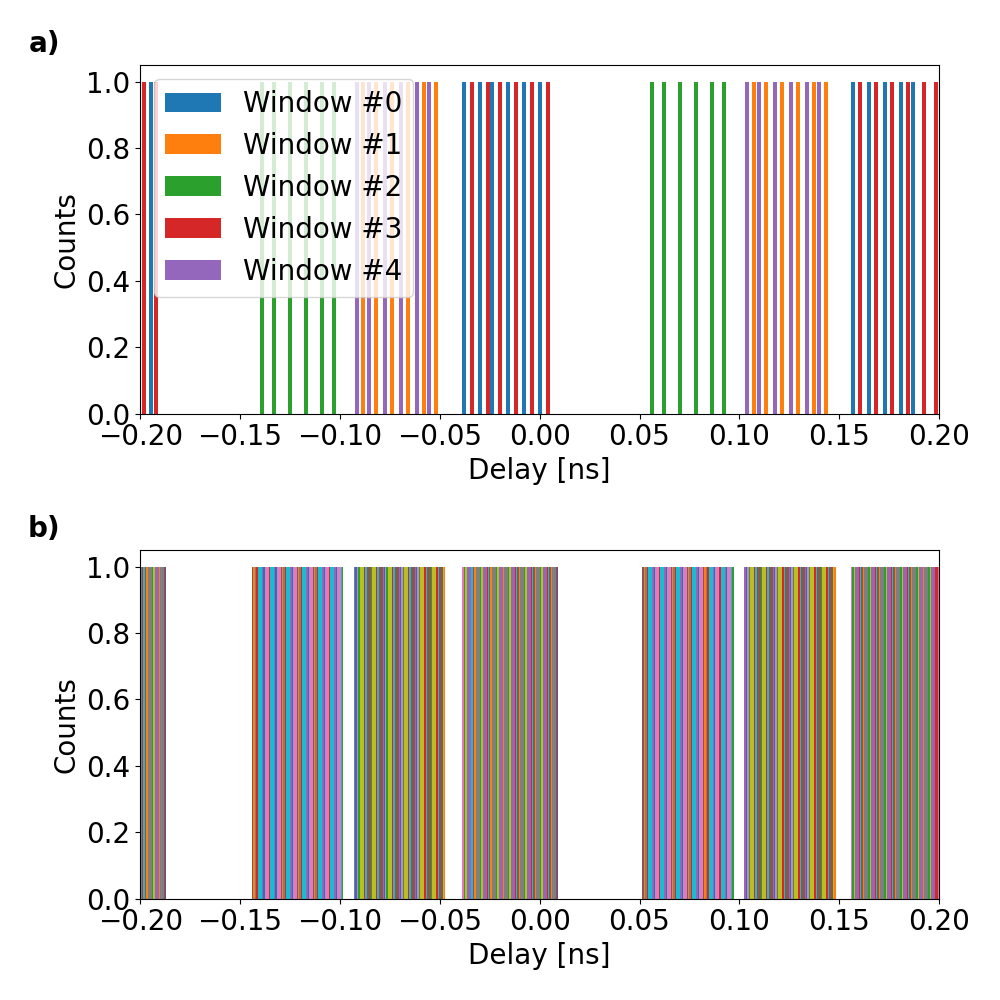}
\caption{Histogram of delay distribution in the interval [-0.1,0.1]~ns for 5 (a) or 150 (b) subsequent shooting windows. Each different color corresponds to a different shooting window.}
\label{study_shooting_windows}
\end{figure}

The time delays (a) and their histogram (b) show a distribution peaked over three sets of achievable delays separated by two forbidden zones. This is peculiar of the example under study and is due to the combination of the three involved frequencies, $f_1$, $f_2$ and $f_{trig}$, in particular the fact that, as described in Eq.~\ref{histo_shift}, every new shooting window shifts by 145$\pm40$ps. As shown in Fig.~\ref{study_shooting_windows}a, where the first 5 shooting windows are reported in different colors over the interval [-.1,.1]~ns, starting from shooting window \#3 the accumulated shift is 394~ps. Since this number is approximately multiple of 195, every third window almost the same set of available delays will be reproduced, generating an overall pattern of permitted and forbidden zones. This is clearly illustrated in Fig.~\ref{study_shooting_windows}b, where 150 windows are reported.

Going back to Fig.~\ref{fig_ampli}, we observe that the simulation results agree with the conclusions of previous section. With regards to the delay distribution, the achieved delays are between roughly $\pm 60$~ps: this is consistent with the expected limitation of $\pm77.5$~ps (Eq.~\ref{aiming_res}) and the higher precision is a consequence of the peculiar histogram pattern. It is interesting also to observe that the width of each set of achieved delay is $8$~ps (Eq.~\ref{pick_res}), which is also a consequence of the histogram pattern and suggests that for each shooting window there is a gray line crossing at least one of the three delay values at roughly -.5, 0 and .5~ns.

The gray plots of Fig.~\ref{fig_ampli}c and \ref{fig_ampli}d are the results of the same simulation discussed so far, with the addition of a small noise of $10^{-11}/$~ms for the ratio $f_1/f_2$, which is in agreement with the experimental measurements.
We observe that the delay distribution becomes more homogeneous in a $\pm77.5$~ps interval, as predicted by Eq.~\ref{aiming_res}. An overdensity of achieved delays is still observed at zero with a peak width of $8$~ps, corresponding to the distribution predicted in case the target is on a gray line (Eq.~\ref{pick_res}). 

\subsubsection{Comparison with experimental data}
\label{Comparison_ampli}
In order to validate our simulations we used two amplified systems: a Quantronix Integra C and a Coherent Libra-HE, equipped with the two oscillators presented above. We connected them to the ADASOPS device and recorded the delays of the amplified pulses, with the same setup as in Figure \ref{setup}.
The oscillator frequencies were the same as those used for the simulations of Case \#2.
The comparison between simulation results and experimental data - gray and black plots of Fig.~\ref{fig_ampli}c and \ref{fig_ampli}d - shows very good agreement. In both cases there is a lower density of positive delays close to zero, which is a residue of the peak distribution observable without noise. Proceeding with longer acquisitions or simulations, the density hole does not fill, but shifts, oscillating between a position on the positive axis to a roughly symmetrical position on the negative axis as the ratio $f_1/f_2$ oscillates. 

\begin{figure}[!ht]
\centering
\includegraphics[scale=.5]{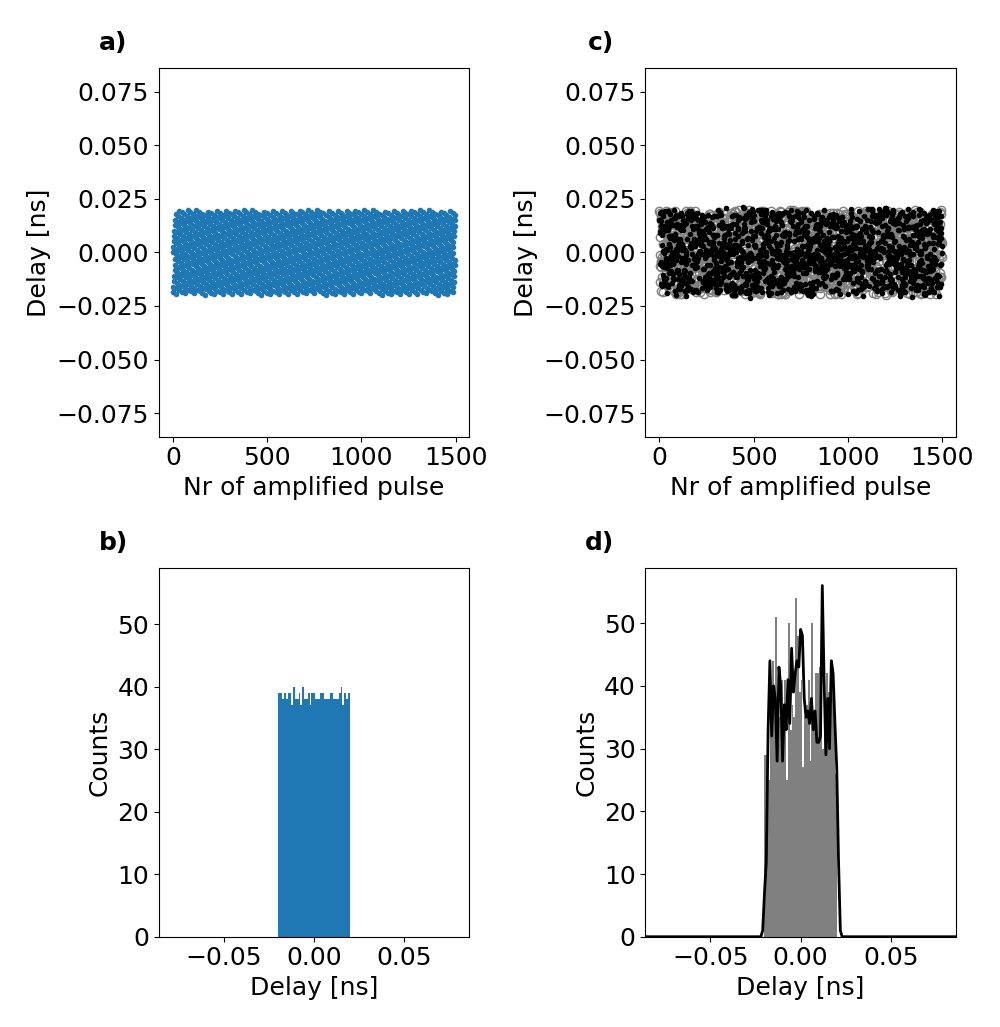}
\caption{Delay distribution for 1500 amplified pulse pairs for two laser systems having a small relative change of frequency of $1.7\times10^{-5}$ compared to the systems of Fig.~\ref{fig_ampli}: (a) Simulation of delay vs. pulse number without noise, (b) histogram of delay distribution without noise. In (c) and (d) the gray plots show the same quantities as in (a) and (b) but taking noise into account, while the black plots are experimental data. }
\label{fig_ampli_newTest}
\end{figure}

It must be pointed out that the aiming resolution is very sensitive to the actual frequencies of the oscillators. As an example, Fig.~\ref{fig_ampli_newTest} shows the result obtained with slightly different oscillator frequencies, namely $f_1^\prime = 29,981,355$~Hz and $f_2^\prime = 80,107,909$~Hz (Case \#2' in Table \ref{TableAmpli}). Despite the tiny frequency shift, i.e. a relative change of $1.7\times10^{-5}$ following the servicing of laser 1, the delay distribution is now considerably narrower. This can be easily interpreted by considering the new frequency ratio. While the first five convergents are identical, the sixth convergent is 317/847 instead of 1597/4267 with the previous frequencies. With the new frequencies, we are now in a nearly ideal situation as the denominator $D_6 = 847$ is just below the number of pulses $N = 961$ in the shooting window. According to eq.~\ref{aiming_res} the aiming resolution is expected to be better than $T_1/D_6 \approx 39$~ps, in excellent agreement with Fig.~\ref{fig_ampli_newTest}.

\section{Correlations between delay, elapsed time and laser energy}

\subsection{Elapsed time from the previous amplified pulse}

In section \ref{delay_in_shoot_win} we introduced the concept of elapsed time $\mathscr{T}$ for an amplified pulse with respect to the previous amplified pulse. Because of the working principle of kHz-ADASOPS, $\mathscr{T}$ is continuously adjusted to obtain the requested target delay. Observations of Fig.~\ref{shooting_window}b revealed a correlation of $\mathscr{T}$ with the gap between the realized delay and the target. The correlation can be direct or inverse, depending on the system.

\begin{figure}[!ht]
\centering
\includegraphics[scale=.5]{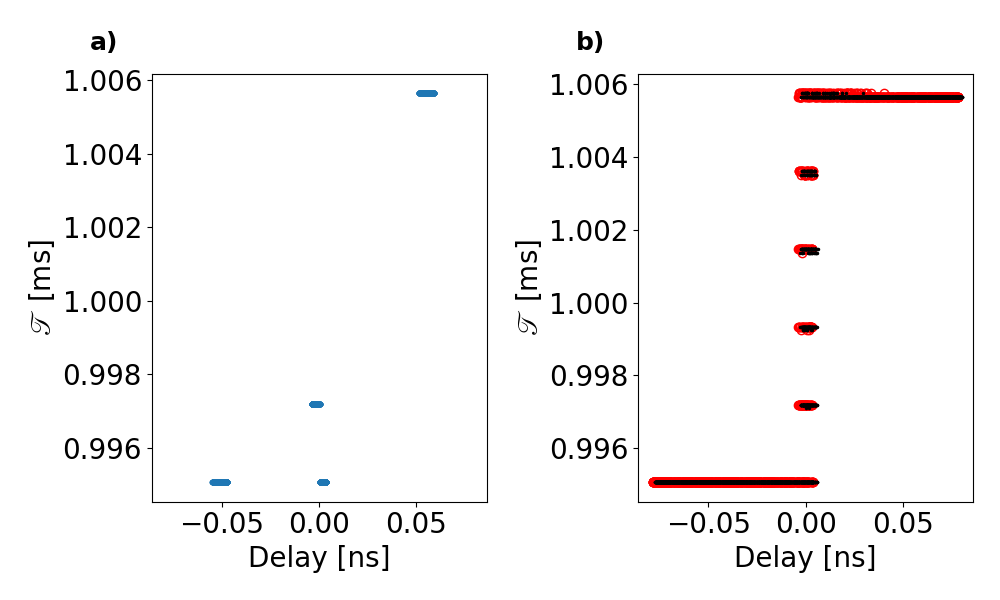}
\caption{Simulation of elapsed time for two laser systems having $f_1$=29,981,872~Hz, $f_2$=80,108,101~Hz (Case \#2 in Table \ref{TableOsc} and \ref{TableAmpli}) and amplifier frequency $f_{trig}=1$~kHz: (a) results without noise, (b) the red plots are results of the simulation with noise and the black plots are experimental data.}
\label{fig_elapsedTest57}
\end{figure}
In Fig.~\ref{fig_elapsedTest57}a are reported the results of the simulation without noise presented in section \ref{simu:data} in terms of elapsed time $\mathscr{T}$ vs. achieved delay. As expected, there is a strong correlation between these two variables. In particular, all the pulses of the set centered at negative delays have $\mathscr{T}=0.995$~ms: knowing the negative slope of the gray lines from Fig.~\ref{shooting_window}b, they are at the beginning of the shooting window and correspond to the first available delay on the grey line closest to the target, which is above it.
Analogously, all the delays of the set centered at positive values have $\mathscr{T}=1.0056$~ms and come from the last available pulse of the shooting window, among the ones on the grey line closest to the goal, which is below it.
On the other hand, all the pulses giving rise to delays in the zero centered set, come from the particular situation where the target is on a gray line. For this reason the correlation is locally inverted, all the delays higher than the target being associated to lower numbers of L$_2$ pulses and vice versa. More precisely we can also say that in this particular case the target is always in between the first and the second available delays of the gray line. 

When taking into account the noise, Fig.~\ref{fig_elapsedTest57}b, we observe an excellent agreement between data (black) and simulation (red). Most of the shots are at the edges of the shooting window, with $\mathscr{T}=0.995$~ms or $\mathscr{T}=1.0056$~ms,  because the target is statistically very often found outside the gray lines. In contrast, the shots closer to zero are those with the target on the grey line: the inverse correlation of this particular case, that was highlighted in the case without noise, is blurred now by the frequency noise. The distribution in six groups of measurements comes from the six available values of delays on the gray line (Eq.~\ref{n_max}) and from the fact that now the target falls with the same probability between any pair of these delays.

As already observed, changing the frequencies of the laser systems, even slightly, will change the pattern of the delay distribution and the type of correlation between delay and $\mathscr{T}$, but the equations \ref{k_min} to \ref{pick_res} will hold. 

The laser systems used for the experiments have regenerative amplifiers pumped by Q-switched lasers (GMP Darwin-527-20M and Evolution-30). They are based on Nd:YLF rods continuously pumped by laser diodes. In this configuration the energy stored into the crystal and the output power are related to the time between two openings of the Q-switch \cite{Paschottaq_switching, PONTIGGIA1978}. In other words, variations of time laps between two consecutive amplified pulses can be expected to have an effect on the energy of the second pulse.

\begin{figure}[!ht]
\centering
\includegraphics[scale=.7]{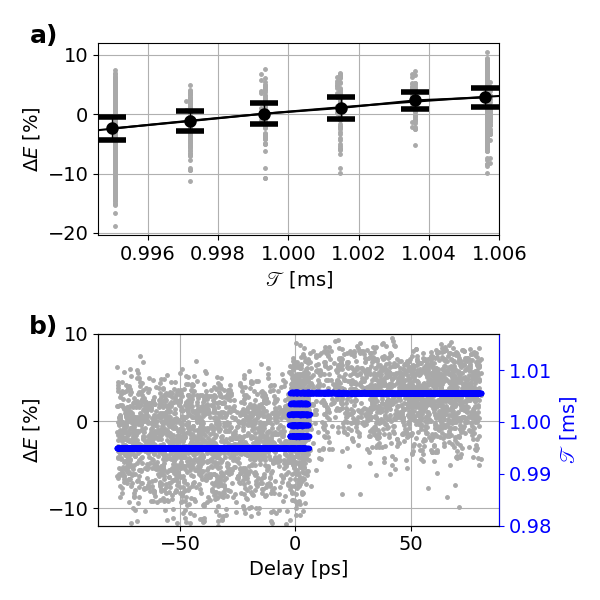}
\caption{a) Correlation between spectral energy variation and elapsed time with respect to the last amplified pulse. In gray are the measurements and in black are the averages and error bars calculated for each sets of $\Delta E$ with similar $\mathscr{T}$. The black line is a linear fit. b) Energy variation (gray) and elapsed time variation (blue) as function of achieved delay. }
\label{correlations}
\end{figure}

Using the setup represented in Figure~\ref{setup_pump_probe} without sample, we recorded the spectra of the L$_2$ pulses via a CCD camera and integrated them to calculate the pulse-to-pulse spectral energy. In Fig.~\ref{correlations}a is reported the energy corresponding to each pulse versus its elapsed time $\mathscr{T}$ with gray dots. Predictably, we find the same distribution in six groups. The black circles are the average energy measured for each group and the black line is a linear fit showing that the relation between energy and $\mathscr{T}$ can be approximated as directly linearly proportional. The energy dispersion is roughly $3\%$~RMS for each group - the relatively high dispersion is due to the fact that the L$_{2}$ pulses are subjected to a series of nonlinear optical processes before reaching the CCD and, in order to highlight the instabilities under study, the non-linear setup was deliberately poorly aligned.

In agreement with the correlation between $\mathscr{T}$  and the achieved delay, demonstrated in Fig.~\ref{fig_elapsedTest57}, a similar type of dependency can be observed between the delay and the energy in Fig.~\ref{correlations}b (gray dots). It is important to stress that, while the energy always varies directly with $\mathscr{T}$, the correlation energy vs. delay depends on the correlation $\mathscr{T}$ vs. delay, which is related to the slope of the gray lines and, as we stated, depends on the ratio $f_1/f_2$.

To sum up, the theoretical comprehension of the process of pulse selection in kHz-ADASOPS has shown that the method intrinsically produces a correlation between the elapsed time between amplified pulses and the gap between the target and the achieved delay. The correlation can be direct or inverse depending on the ratio between oscillator frequencies. At the same time, in the context of our experiments, the Q-Switch technology in amplified systems always generates a direct correlation between the elapsed time and the pulse energy. Therefore, there is always a correlation, direct or inverse depending on the oscillator frequencies, between the energy and the delay gap. This correlation can be problematic in ADASOPS driven experiments because it can generate artifacts, particularly with regards to the baseline as previously reported\cite{Helbing2023}. However, the following section reports a method to compensate for the possible artifacts and discusses several approaches to overcome this drawback.

\subsection{Impact on probe-energy fluctuations}
\label{impact on probe-energy fluctuations}


Using the two amplified laser systems presented above, we now apply the ADASOPS method on a visible pump - mid-infrared probe experiment in order to record the dynamic response of heme-bound carbon monoxide molecule in carboxy-hemoglobin. While thorough results have been reported earlier~\cite{Nuernberger2010}, we focus here on the impact of ADASOPS on noise.
The visible pump pulse is obtained by frequency doubling the output of the Integra into a BBO crystal in order to obtain a 400-$nJ$, 390-$nm$, 100-$fs$ pulse. The probe pulse, generated by the Libra, undergoes Optical Parametric Amplification (OPA) and Difference Frequency Mixing to generate a mid-infrared spectrum centered at 1950 cm$^{-1}$. After going through the sample, the beam is up-converted~\cite{Lee09} and directed to a visible CCD camera (Princeton).
While the Libra repetition rate is 1~kHz, the Integra is operated at 500~Hz so as to record a reference pump-off signal ($\mathscr{R}$) every pump-on acquisition ($\mathscr{S}$). The transmittance variation is calculated as:

\begin{equation}
  \frac{\Delta T}{T} = \frac{\overline{\mathscr{S}-\mathscr{R}}}{\overline{\mathscr{R}}},
  \label{deltaTsurT}
\end{equation}

where $\mathscr{S}$ and $\mathscr{R}$ are consecutive acquisitions and the over-line symbol represents the average. 

\begin{figure}[!ht]
\centering
\includegraphics[scale=.5]{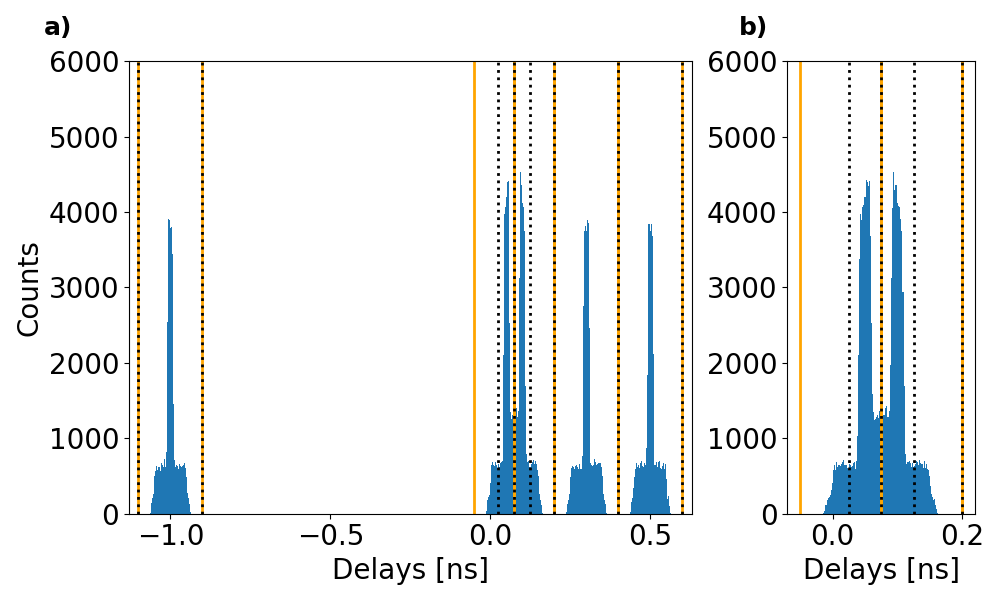}
\caption{Histogram of achieved delays. a) Graph reporting all the achieved delays aiming continuously at -1, 0.05, 0.1, 0.3, and 0.5~ns. The particular distribution around each target delay is consistent with the simulation discussed in section \ref{simu:data}. The orange lines indicate the bin edges according to the basic ADASOPS algorithm and the resolution parameter set at 100~ps. The black dotted lines are the bin edges in case of compensation algorithm, with the same resolution parameter. b) Zoom in the interval [-0.07, 0.22]~ns where the distributions of the delays obtained aiming at 0.05 and 0.1 overlap. }
\label{histo162}
\end{figure}

\begin{figure}[!ht]
\centering
\includegraphics[scale=.5]{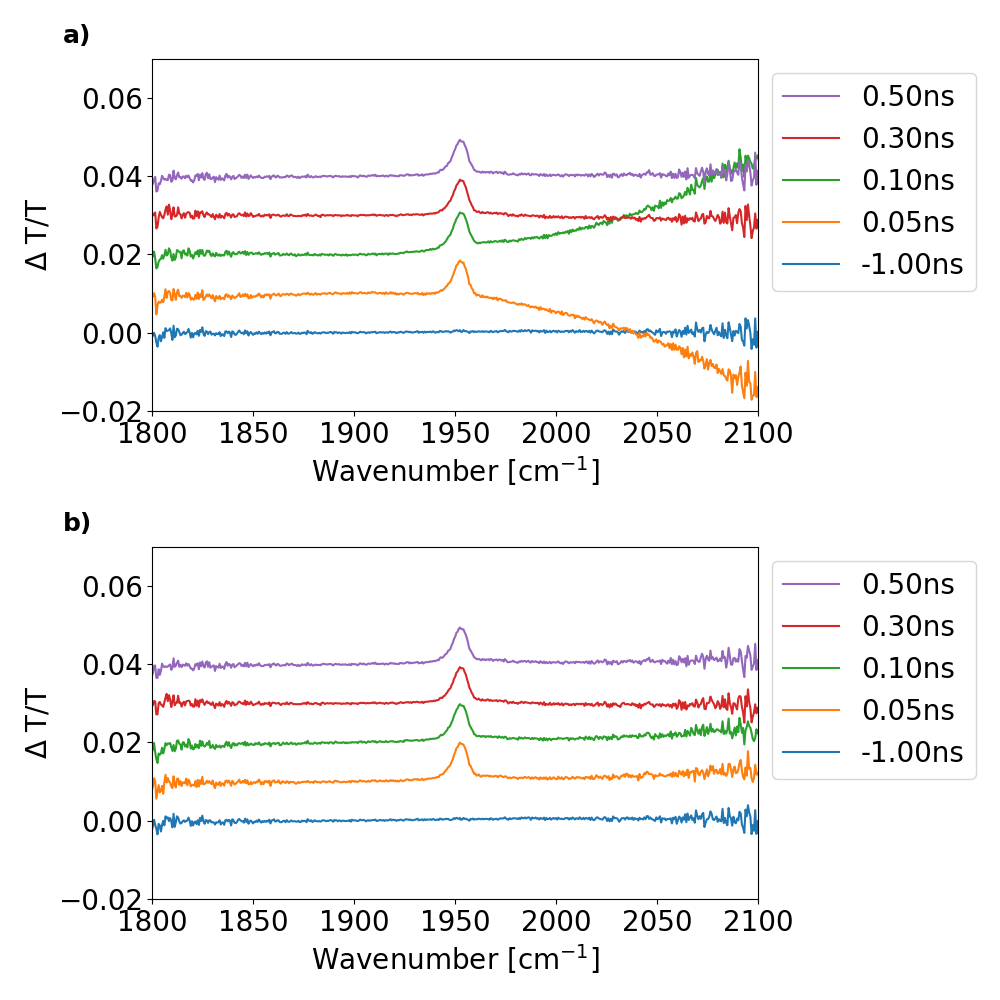}
\caption{Differential mid-infrared spectra of HbCO. a) Straightforward application of ADASOPS method. b) Use of compensation algorithm on the same data as (a).}
\label{HbCO}
\end{figure}

This calculation is performed for each delay bin dividing the delay axis. In practice, the user communicates the requested scanning delays to the ADASOPS device. The device continuously realizes the scan by selecting the pulses with closest delays with respect to the targets. In our case we repeat each delay 100 times, in order to avoid reducing the aiming accuracy as discussed in section \ref{ampli_delays_distribution}. We define a resolution parameter $r$, e.g. 100~ps in this case, that identifies a bin around each target delay so as to take into account only the achieved delays that actually fall into this bin. If there are target delays closer together than the resolution, $r$ is re-assigned as half the interval between these delays (see the bin edges drawn as orange vertical lines in Fig.~\ref{histo162}). Only the pump-on measurements inside a bin are used to calculate the $\Delta T/T$ relative to the corresponding target delay, using the next pump-off measurements as reference. 
In Fig.~\ref{HbCO}a are reported the $\Delta T/T$, vertically shifted by a gap of 0.01 for clarity, for the following target delays:  -1, 0.05, 0.1, 0.3 and 0.5~ns. We can see the ultrafast photobleach of the bound state as a peak at 1951~cm$^{-1}$ appearing before 0.05~ns. The orange plot (0.05~ns) and the green plot (0.1~ns) show a significant artifact as a bend in the baselines. This phenomenon is related to the delay vs. energy correlation, as the combination of energy fluctuation and nonlinear conversion results in a distortion of the mid-infrared spectrum. If we look at the histogram of achieved delays in Fig.~\ref{histo162}, for each target we find a flat distribution roughly 100~ps wide with a narrow peak in the centre, consistently with the simulations described in section \ref{simu:data} and with the laser frequencies measured during acquisition: $f_1$=29,981,639~Hz and $f_2$=80,107,871~Hz (Case \#2'' in Table \ref{TableAmpli}).
We observe that the bins for -1, 0.3, and 0.5~ns are large and therefore all the achieved delays aiming the target are inside the bin: the distribution is symmetrical, compared to the target in the centre, so that the correlation delay-energy is compensated by the average calculations of $\Delta T/T$.  However, for the bins at 0.05 and 0.1~ns, not only the peak is off-centre, but there is also an overdensity of delays on the side of the nearest target (see the magnification in Fig.~\ref{histo162}b). In fact some of these delays come by aiming at the slightly smaller target - and have slightly smaller energy -, while others come by aiming at the slightly larger one - and have slightly higher energy. Therefore, in these overdensities the correlation of the energy with respect to the bin delay can be direct or indirect, depending on what the aimed target was. 
In these bins the overall effects cannot compensate and a residual of the delay-energy correlation remains and has opposite effect in 0.05 and 0.1~ns (orange line and green line in Fig.~\ref{HbCO}a) because the overdensity is on opposite sides with respect to the target.

As expected, the correlation delay gap vs. $\Delta E$ can generate experimental artifacts. Inside each delay bin the energy variation evolves monotonically: if the bin is centered with respect to the target delay, the variations will be averaged out, otherwise they will result in an artificial signal.

\subsection{Compensation methods}

The theoretical understanding of the artifacts and their origins makes it possible to compensate for them in various ways, depending on the type of experiment.
In post-treatment analysis for example it is possible to ensure symmetrical bins (such as the dotted black lines in Fig.~\ref{histo162}) and exclude all the pulses aiming at different targets, although falling into the bin.
In Fig.~\ref{HbCO}b the previous data are treated with a compensation algorithm and the artifacts have successfully disappeared. Nevertheless, this method reduces the number of measurements, and inhomogeneously in relation to the delays, causing lower signal to noise ratio for narrower bins.

Other methods can also be considered, in particular using the reference signal $\mathscr{R}$ to automatically compensate for artifacts. Many pump-probe experiments are based on setups including a simultaneous acquisition of signal and reference provided by the same laser pulse. In this case the average calculations intrinsically cancel out the effects of energy variations.

\begin{figure}[!ht]
\centering
\includegraphics[scale=.7]{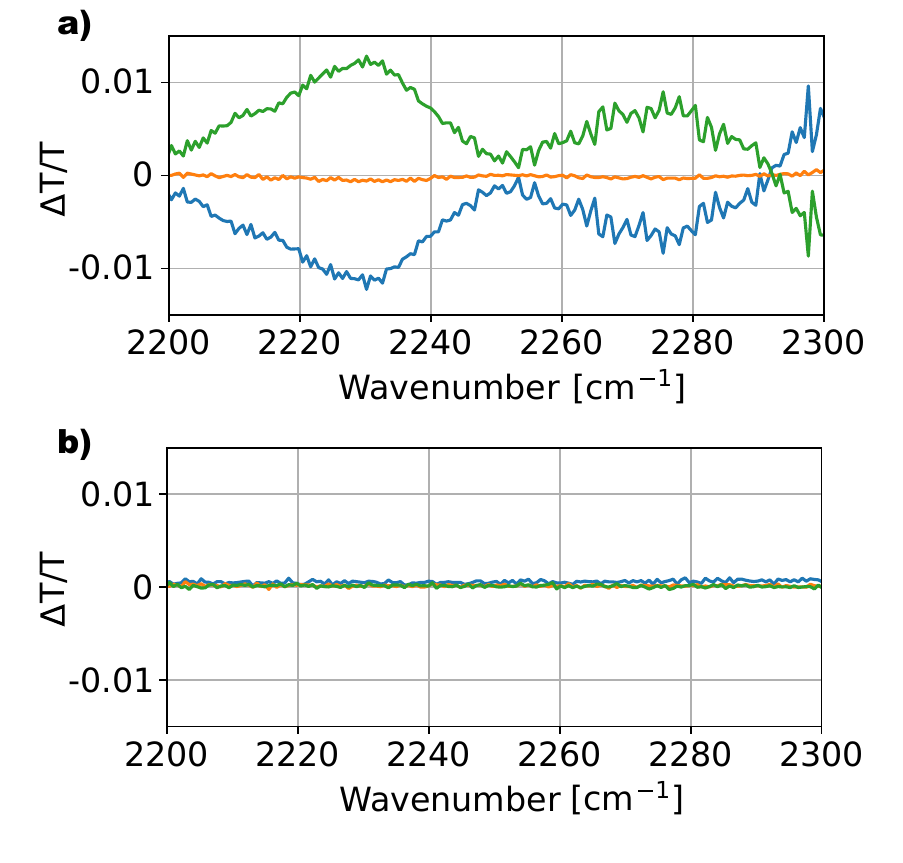}
\caption{Differential mid-infrared spectra acquired without sample for noise characterization. a) Measurement artifacts due to laser energy fluctuations associated with ADASOPS. b) Absence of artifact in ADASOPS measurements performed with pairing algorithm. Different colors correspond to different selection criteria, as explained in the text.}
\label{pairing_effect}
\end{figure}
In situations where the reference $\mathscr{R}$ comes from another pulse than the signal $\mathscr{S}$, as in our experiment, or e.g. when the pump is chopped, the compensation algorithm can be integrated in the process of pulse selection, simply by imposing the same elapsed time on the reference pulses as on the previous signal pulses. This is possible because the probe pulse, used to record a reference, is not coupled with a pump pulse,  hence its delay has no physical meaning and can be freely adjusted.
We name this method "pairing algorithm" because the probe beam becomes a series of probe-reference pulse pairs having the same elapsed time, and so the same energy. This algorithm is coded into the ADASOPS device and can be straightforwardly activated just by selecting the specific operational mode.
As a demonstration of its high efficiency, we repeated similar experiments as described in \ref{impact on probe-energy fluctuations}, but removing the sample, in order to get a direct measurement of the experimental noise.  
In Fig.~\ref{pairing_effect} are reported a series of measurements recorded while targeting the same delay. The data are classified in three exclusive subsets with respect to the distribution of the achieved delays: the orange curves are associated with delays centered around the target, whereas the blue and the green curves are obtained with delays respectively below and above the first set. Fig.~\ref{pairing_effect}a has been recorded without compensation algorithm. The difference between the blue and the green curves is an indication of symmetric modifications in the signal spectra for delays exceeding or falling short of the target, which can be attributed to fluctuations in laser energy. In contrast, the low noise observed in the orange curve results from the mitigation of these effects for delays centered on the target. On the other hand, Fig.~\ref{pairing_effect}b has been recorded by switching on the pairing algorithm: its effectiveness is illustrated by the disappearance of the large modulations for the blue and the green curves and the independence of the noise with respect to the subset of origin.

This pairing method, described in detail for the first time in the present article, has already been implemented in previous experiments \cite{Aleksandrov2024}. 

\section{Conclusions and perspectives}

We provided a comprehensive theoretical description of delay evolution in asynchronous optical scanning, based on the ratio between laser frequencies expressed in its continued fraction form. 
Knowing the oscillator frequencies with Hz precision, the delay distribution and the theoretical ADASOPS resolution limit are determined with sub-picosecond accuracy.
By extending this model to the selection algorithm applied in kHz-ADASOPS for the choice of amplified pulses, we can also predict the aiming accuracy. 

We demonstrated a correlation between achieved delays and elapsed time between two amplified pulses, implying a correlation delays vs. energy, resulting in energy instabilities and possible artifacts.
Although these artifacts can be limited by careful experimental alignment, we also studied methods of compensating for these artifacts, either algorithmically or through the use of a pairing approach.

The ADASOPS method is now a well understood and controlled method for time-resolved spectroscopy over extremely wide time intervals, extending from hundreds of femtoseconds to milliseconds. Electronic delay control allows for rapid and flexible scanning without the limitations due to moving parts, typical of more conventional methods. This makes the ADASOPS method ideally suited for accessing broader time scales. However, the sub-picosecond resolution has to be improved in order to access even faster processes, as coherent interaction of photons and/or charges, in the time-range of hundreds of femtoseconds. A recent application of kHz-ADASOPS to dual-comb spectroscopy \cite{Baltuska23} demonstrated a resolution of 80~fs, and suggested the possibility to go as low as 1~fs.
Remaining on standard femtosecond laser systems we plan to improve ADASOPS resolution by combining the method with Fourier-Transform Spectral Interferometry (FTSI) \cite{Joffre95, Joffre00} for accurate measurement of delays shorter than 30~ps, as previously anticipated \cite{Joffre_17_oe, Baltuska23}. Using spectral interference between two leaks of the pump and probe beams - resorting to continuum generation in case their spectra do not overlap -, we will extract the spectral phase and calculate sub-ps pump-probe delays with a final resolution expected to be only limited by the pulse duration.

\bibliographystyle{apsrev4-1}
\bibliography{bibliography.bib}

\newpage

\begin{figure}[!ht]
\title{TOC Graphic}

\centering
\includegraphics[scale=1]{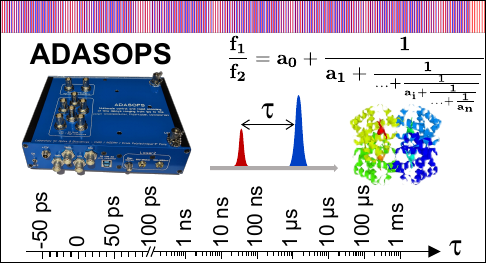}
\label{TOC Graphic}
\end{figure}
\end{document}